\documentclass[sigconf, nonacm]{acmart}

\AtBeginDocument{%
  }

\setcopyright{acmlicensed}
\copyrightyear{2018}
\acmYear{2018}
\acmDOI{XXXXXXX.XXXXXXX}

\acmConference[Conference acronym 'XX]{Make sure to enter the correct
  conference title from your rights confirmation emai}{June 03--05,
  2018}{Woodstock, NY}
\acmISBN{978-1-4503-XXXX-X/18/06}

\begin{document}

\title{Agentic Workflows for Conversational Human-AI Interaction Design}

\author{Arthur Caetano}
\email{caetano@ucsb.edu}
\orcid{0000-0003-0207-5471}
\affiliation{%
  \institution{University of California}
  \city{Santa Barbara}
  \state{CA}
  \country{USA}
}

\author{Kavya Verma}
\affiliation{%
  \institution{University of California}
  \city{Santa Barbara}
  \state{CA}
  \country{USA}
}

\author{Atieh Taheri}
\affiliation{%
  \institution{Carnegie Mellon University}
  \city{Pittsburgh}
  \state{PA}
  \country{USA}
}

\author{Radha Kumaran}
\affiliation{%
  \institution{University of California}
  \city{Santa Barbara}
  \state{CA}
  \country{USA}
}

\author{Zichen Chen}
\affiliation{%
  \institution{University of California}
  \city{Santa Barbara}
  \state{CA}
  \country{USA}
}

\author{Jiaao Chen}
\affiliation{%
  \institution{Georgia Institute of Technology}
  \city{Atlanta}
  \state{GA}
  \country{USA}
}

\author{Tobias Höllerer}
\affiliation{%
  \institution{University of California}
  \city{Santa Barbara}
  \state{CA}
  \country{USA}
}

\author{Misha Sra}
\affiliation{%
  \institution{University of California}
  \city{Santa Barbara}
  \state{CA}
  \country{USA}
}

\renewcommand{\shortauthors}{Caetano et al.}

\begin{abstract}
Conversational human-AI interaction (CHAI) have recently driven mainstream adoption of AI. However, CHAI poses two key challenges for designers and researchers: users frequently have ambiguous goals and an incomplete understanding of AI functionalities, and the interactions are brief and transient, limiting opportunities for sustained engagement with users. AI agents can help address these challenges by suggesting contextually relevant prompts, by standing in for users during early design testing, and by helping users better articulate their goals. Guided by research-through-design, we explored agentic AI workflows through the development and testing of a probe over four iterations with 10 users. We present our findings through an annotated portfolio of design artifacts, and through thematic analysis of user experiences, offering solutions to the problems of ambiguity and transient in CHAI. Furthermore, we examine the limitations and possibilities of these AI agent workflows, suggesting that similar collaborative approaches between humans and AI could benefit other areas of design.
\end{abstract}


\begin{teaserfigure}
    \centering
  \includegraphics[width=.8\textwidth]{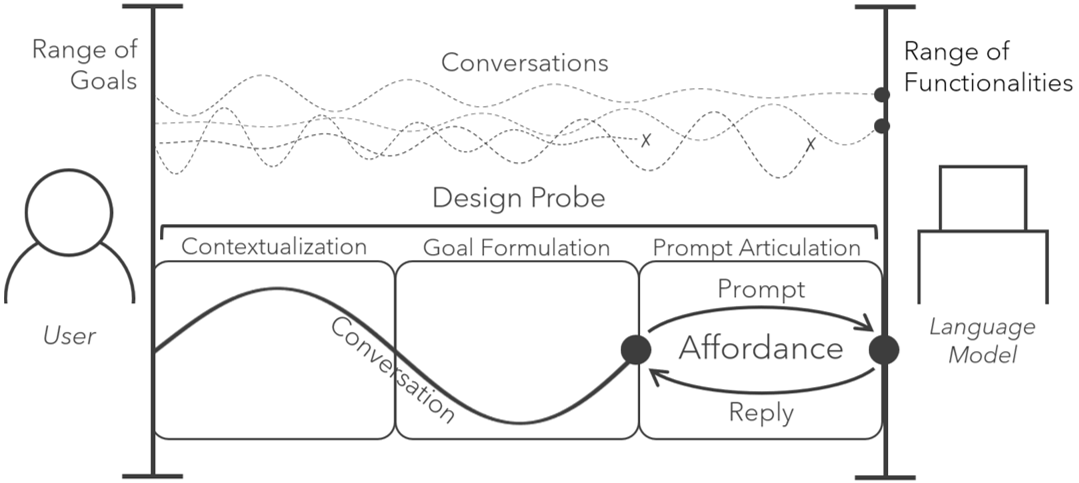}
  \caption{Conversational human-AI interactions can be understood as dialogue exchanges between user goals and AI functionalities. When a user goal is articulated as a prompt and an AI functionality produces a satisfactory result, an affordance is established. We consider two main challenges to this dynamic: ambiguity and transience. The breadth of user goals and AI functionalities introduces ambiguity. The short-lived and seldom-repeated nature of these interactions adds the challenge of transience. Our design probe proposes a workflow composed to guide human-AI conversations through the steps of contextualization, goal formulation, and prompt articulation. In each of these steps, humans are assisted by agents such as User Proxies, Contextual Personas and Goal Refinement agents.}
  \Description{}
  \label{fig:teaser}
\end{teaserfigure}


\maketitle

\section{Introduction}
Artificial intelligence (AI) systems powered by large language models (LLMs) have transformed creativity and productivity applications~\cite{vaccaro2024combinations}.
Conversational human-AI interaction (CHAI) is the prevalent paradigm in these systems, and has been crucial to their widespread adoption~\footnote{\url{http://chat.com}, \url{http://claude.ai}}~\cite{zheng2022ux}. CHAI uses conversational user interfaces (CUIs) to leverage the linguistic capabilities of AI and allow users to express goals in natural language to access AI functionalities through dialogue. Active research areas in CHAI include modeling user behavior in conversational interactions~\cite{subramonyam2024bridging, zamfirescu2023johnny} and proposing domain-specific interfaces that complement dialog~\cite{masson2024directgpt, liu2023wants}. Although important progress has happened in this direction, several obstacles hinder CHAI research and design.
The complexities of AI as a design material~\cite{yildirim2022experienced, feng2023ux, yildirim2023creating, yildirim2023investigating} and the inherent limitations of conversational interfaces~\cite{fischer2019progressivity, zargham2022understanding, skjuve2023user, norman2010natural} challenge designers and users alike. We consider two properties of conversational interactions with AI to be at the core of these challenges: ambiguity and transience.

\textbf{Ambiguity} in CHAI emerges from the superposition of the wide range of functionalities of conversational AI systems and the amplitude of possible user goals they can support. This combination creates a vast design space, where conversations between users and AI can be imagined as trajectories going from a user goal to an AI functionality, as depicted in Figure~\ref{fig:teaser}. When the user is able to conduct a dialogue with AI that satisfies their goal, an affordance is established~\cite{norman2013design}. Because, there are many (or no) such dialogues, the affordances of the systems became ambiguous, a phenomenon called  the ``capability gap''~\cite{subramonyam2024bridging}. The limitations of CUIs in presenting signifiers for affordances~\cite{norman2010natural, hutchins1985direct, masson2024directgpt} aggravates the problem. In this setting, designers act as facilitators of human-AI dialogue, using design interventions to guide users through the design space. However, it is difficult to predict users' goals before they interact with the system, limiting opportunities for upfront design.

\textbf{Transience} is another challenging characteristic of conversational interactions with AI. These interactions are typically driven by immediate, momentary needs and are often resolved through brief exchanges that users rarely revisit. Designing for such spontaneous and ephemeral interactions presents unique challenges, as we cannot rely on prolonged user engagement or iterative refinement, the cornerstones of user-centered design. Employing user-led generative design is a promising alternative~\cite{heer2019agency}, but transferring large portions of the design process to users can create friction and harm user experience by obstructing the direct pursuit of their initial goal.

As presented, navigating ambiguity and transience in CHAI requires context-specific solutions for users as well as replicable practices for designers. To investigate these solutions, we employed a Research-through-Design (RtD) approach ~\cite{zimmerman2007research, zimmerman2010analysis, gaver2012should} because it supports iterative exploration of directions emerging~\cite{gaver2022emergence} from engagement with users and the design practice, all desirable methodological qualities for our goal. As research instrument, we used a design probe~\cite{boehner2007hci} in the form of an AI chat web app powered by a small language model~\cite{abdin2024phi}, iteratively designed with 16 users over four design cycles. Our probe explores the technical opportunities offered by agentic workflows with humans-in-the-loop---structured sequences of activities that involve collaboration and decision-making among humans (users and designers) and AI agents, where each participant has distinct roles and responsibilities within a process. In summary, our RtD inquiry is: \textbf{How can agentic workflows help users and designers navigate ambiguity and transience in CHAI?}

Our design probe implements a structured workflow consisting of three stages: contextualization, goal formulation, and prompt articulation. In the contextualization stage, users provide relevant information with minimal effort, facilitated by automated API calls and image uploads. During goal formulation, users receive personalized recommendations from agents tailored to their persona and contextual data, with the option to select these recommendations or add their own goals. A goal refinement agent assists in refining these goals iteratively, merging similar ones and breaking down complex goals into smaller, actionable components. Finally, in the prompt articulation stage, agents generate tailored recommendations based on the finalized goals and user context. In its fourth iteration, the probe incorporates a designer-centered workflow, allowing designers to interact with app usage data through conversations with a User Proxy agent, enhancing their ability to explore and understand user interactions more deeply.

As a comprehensive account of our design probe, we present an annotated portfolio~\cite{bowers2012logic, gaver2012should} detailing the workflows across the four versions, highlighting their limitations, user reception, and potential to mitigate ambiguity and transience. Moreover, we share insights from user engagement drawn from quantitative analysis of probe usage and reflexive thematic analysis~\cite{braun2024deductive, clarke2013successful} of user feedback data, narrating the overall user experience with the probes. Lastly, we discuss implications of our workflows and the use of agents in CHAI design. Our contributions are:

\begin{enumerate}
    \item An annotated portfolio of a design probe using agentic workflows with humans-in-the-loop to navigate ambiguity and transience in CHAI.
    \item An account of user experience with our probe in the form of themes and analysis of the range of goals and functionalities found with our probe.
    \item The final version of our design probe as an artifact implementing the workflows.~\footnote{Code, prompts, and supporting are available at \url{https://github.upon.publication}.}
\end{enumerate}

From our RtD exploration with a design probe, we identified agentic workflows with humans-in-the-loop that can help designers and users to navigate ambiguity and transience in CHAI. These findings can benefit users by helping them to clarify intentions and articulate effective prompts to access AI affordances. We expect that designers adopting these workflows will benefit from enhanced ability to find user needs in conversational AI, creating new opportunities for design.

\section{Related Work}

\subsection{Conversational Human-AI Interaction}



Conversational User Interfaces (CUIs) are systems that enable dialogue-based interaction that mimics human conversation, and are widely in use today in the form of chatbots~\cite{folstad2017chatbots} and voice activated virtual assistants~\cite{perez2018everybody}, among others. CUIs have shown promise in applications to health~\cite{kocielnik2021can, vaidyam2019chatbots}, education~\cite{haristiani2019artificial, colace2018chatbot} and customer service~\cite{gupta2015commerce, Johannsen2018ComparisonOC}, in addition to their extensive use as intelligent personal assistants~\cite{de2020intelligent} such as Amazon Alexa and Apple's Siri.

Rapid advances in generative AI and large language models (LLMs)~\cite{achiam2023gpt} and increased access to such software through tools like ChatGPT has made conversational human-AI interaction widely used even among non-experts~\cite{ma2024exploring, humlum2024adoption}. Generative AI and LLM-based tools offer significantly more functional flexibility than previous CUIs, and have been employed for a variety of applications including immersive authoring~\cite{zhang2024vrcopilot}, creative coding~\cite{angert2023spellburst}, and equipping real-world objects with digital functionalities~\cite{dogan2024augmented}, among others. 

This functional flexibility, however, can introduce ambiguity in user interactions with the system. One source of ambiguity is a challenge described as the ``capability gap''~\cite{subramonyam2024bridging}, where the affordances and capabilities of the system are unclear to the user. Making the system's capabilities clear to the user is an established guideline for human-AI interaction~\cite{amershi2019guidelines}, and implementing strategies for discoverability of functionalities in has been shown to improve user experience~\cite{kirschthaler2020can, weng2024insightlens}. Unlike direct manipulation interfaces however, CUIs often lack clear signifiers for AI Affordances~\cite{hutchins1985direct, masson2024directgpt}, and a lack of interpretability in human-AI conversations can also contribute to this challenge~\cite{angert2023spellburst}.   

Another source of ambiguity emerges from inherent ambiguity of natural language, which relies on shared background knowledge as much as semantic content~\cite{brennan1990conversation}. While AI exhibits ``common-sense'' and specialized knowledge~\cite{davis2023benchmarks}, it often lacks realtime awareness of users and their specific contexts---nuances that remain difficult to capture with current multimodal models~\cite{duan2024vlmevalkit}. As a result, users are burdened to explicitly provide context that would typically be assumed in human-to-human conversations, leading to excessive trial-and-error to disambiguate prompts~\cite{zamfirescu2023johnny, liu2023wants, subramonyam2024bridging}.

In addition to the ambiguity of user context and system capabilities, another challenge in CHAI is the transience of interactions. Transience (or ephemerality), which refers to phenomena that are short-lived in timespan, is an inherent characteristic of human life, and has also been integrated into art and architecture~\cite{diller2002blur} and user interface design~\cite{doring2013design}. Human-AI conversation through LLM-based tools is inherently transient, because of rapidly changing user needs and contexts across multiple threads of conversation. This transience in interaction can lead to a lack of user engagement, making the process of designing for CHAI more challenging. Additionally, the challenge of ambiguity in user context is exacerbated with transience, since the system cannot rely on any learned knowledge to disambiguate user goals.

While there had been some progress towards mitigating these issues with design interventions including prompt scaffolding, prompt iteration tracking, and clear signifiers~\cite{zamfirescu2023johnny, angert2023spellburst, subramonyam2024bridging, masson2024directgpt}, there is still limited research on supporting designers of CHAI specifically focusing on the challenges of ambiguity and transience.

\subsection{Agentic Systems}

Agents are systems capable of autonomously perceiving, reasoning, and acting in dynamic environments to achieve specific goals. Advances in LLMs have enabled more powerful agents and allowed users to delegate and instruct them through natural language. Our work explores the technological opportunity of agentic workflows with humans in the loop, i.e., workflows orchestrates the collaboration between humans and AI agents to achieve a complex goal. This approach offers a balanced mix between automation and agency, as human feedback guides agents in the workflow and the final decisions are taken by humans. Common agent implementations utilize role-play prompting, Chain-of-Thought, and Retrieval-Augmented Generation.

Language models can be prompted to play a role (e.g.``You are [role]''), enabling them modulate their behavior accordingly as they follow subsequent instructions. This technique is known as role-play prompting and has been shown to improve performance in one-shot reasoning~\cite{shanahan2023role, kong2024better}. Prior work has used role-paly prompting to create agents with diverse behaviors and observed emergent behaviors in complex simulations with implications to the design of social platforms~\cite{park2022social, yang2024oasis}. The use of role-playing agents has also grown in the social sciences, as a model of human-like behavior~\cite{dai2024artificial, park2024generative}. Although this technique does not replace user engagement in the design process, it has can serve as a prototyping and computational design method. In a recent study, Shin et al.~\cite{shin2024understanding} proposed AI workflows for generating personas and highlighted their potential to create interactive versions of these personas through role-play prompting~\cite{shin2024understanding}. Our work uses role-play prompting to create persona agents that help users in the prompting process and also serve as interactive proxies of users for designers, helping to tackle ambiguity and transience.

Techniques such as Chain-of-Thought (CoT)~\citep{wei2022chain} prompting enable models to generate intermediate reasoning steps, improving complex problem-solving abilities. However, CoT often generates redundant reasoning steps, which can increase computational overhead and lead to inefficiencies in dynamic tasks~\citep{stechly2024chain}. Moreover, CoT cannot directly link reasoning steps to actionable outcomes, restricting its applicability in environments requiring real-time interactions. The ReAct framework is proposed to address these limitations~\citep{yao2023react}. It integrates reasoning with actionable decisions, mixing thought processes with task execution. Thus, ReAct refines reasoning steps and improving the robustness of outputs in complex environments. ReAct's emphasis on real-time feedback and decision grounding makes it suitable for tasks that requiring external interactions, such as information retrieval and system control~\citep{joshi2024reaper}. In our work users' input and decisions over AI-generated recommendations offer agents external feedback.

While integrating tools into agent frameworks can enhance capabilities, it is not a necessity for effective performance. ~\citep{turtayev2024hacking, shah2024agents} emphasize that agents can operate proficiently without external tools, relying instead on internal mechanisms. ~\citep{zhang2024xlam, hassouna2024llm} show that agents can achieve strong performance in complex tasks without interactive tools, highlighting the sufficiency of inherent reasoning abilities. The integration of tools is a design choice for enhancing the performance, but not a prerequisite for effective agent operation. In this work, agents do not directly act on the environment, but instead interact with users supporting their cognitive process and reducing motor effort during conversations.

The integration of RAG with agent-based frameworks has advanced systems' ability to incorporate external knowledge. Zerhoudi et al.~\cite{DBLP:journals/corr/abs-2407-09394} developed PersonaRAG for user-centric applications that adapts retrieval and generation to user preferences. ~\citeauthor{chang2024main}~\cite{chang2024main} presented MAIN-RAG, a collaborative agent system for filtering and ranking retrieved documents. Our work utilizes RAG techniques to extract contextually relevant persona descriptions from a publicly available dataset~\cite{ge2024scaling}. These personas are role-played by agents that suggest prompts and goals to users.

\section{Methodology}

To explore how users and designers can leverage agentic workflows to navigate ambiguity and transience in CHAI, we lean on Research-through-Design (RtD)~\cite{zimmerman2007research, zimmerman2010analysis, gaver2012should}. RtD supports iterative exploration of emergent~\cite{gaver2022emergence} directions introduced by the engagement with users and the design practice, crucial methodological characteristics to investigate our design inquiry. The instrument of our investigation is a probe consisting of a AI chat web app powered by a small language model~\cite{abdin2024phi} that explores the technical opportunities of agentic workflows with humans-in-the-loop. The probe was iteratively designed with 10 participants over four design cycles, and the process took one month. We collected user feedback, contextual data, and app usage data with participants' informed consent. Our study was approved by our local IRB (\textit{protocol number anonymized}).

\subsection{Materials}

Our design probe was a conversational AI web app developed with Streamlit 1.41~\footnote{\url{https://streamlit.io}}, compatible with major mobile and desktop browsers. The powerhouse of our probe was Microsoft Phi-3.5-vision~\footnote{\url{https://huggingface.co/microsoft/Phi-3.5-vision-instruct}}~\cite{abdin2024phi}, a recent multimodal small language (SLM) model running on our on-premises server~\footnote{Ubuntu 22.04, AMD Ryzen Threadripper PRO 5955WX, 258GB, NVIDIA RTX A6000}. Another key resource for our design probe was Sentence-BERT~\footnote{\url{https://huggingface.co/sentence-transformers/all-MiniLM-L6-v2}}~\cite{reimers2019sentence}, a model that encodes text into 384-dimensional semantic embeddings. Its primary role in our probe was to enable semantic search via cosine similarity on a publicly available dataset of persona descriptions derived from web content~\cite{ge2024scaling}.

Choosing an on-premises SLM allowed us to achieve a trade-off between generative power and sustainability. While SLMs are less powerful than contemporary LLMs, they consume less energy, supporting a more sustainable approach to AI research\cite{strubell2020energy}. Another consequence of this choice was the need to consider model limitations in our design, such as lack of JSON output formatting and system messages~\footnote{\url{https://learn.microsoft.com/azure/ai-studio/how-to/deploy-models-phi-3-5-vision}}, imposing constraints that practitioners may also encounter. Details on how these materials were combined into the design probe are presented in Section~\ref{sec:probe}.

\subsection{Participants}

Our study included 10 participants, each engaging in at least one of four iterations conducted over four weeks. Participants were recruited locally and encouraged to invite others from their communities, leveraging a word-of-mouth strategy. The asynchronous, remote format enabled participation from 18 test locations in total. The sample was diverse in gender (F=5, M=5, NB=0) and age (ranging from 18 to 44 years old). Three participants indicated that they currently have or previously had a disability. Regarding AI experience, 90\% were familiar with ChatGPT, and 63.6\% with Gemini. Additionally, 63.6\% reported using AI multiple times daily for a broad spectrum of tasks including Q\&A, writing and coding assistance, productivity and task management, brainstorming and feedback, academic and professional tasks, as well as creative activities like generating art.

\subsection{Overall Design Process}

After providing informed consent, participants filled out a pre-survey that collected demographic data, perception and usage of AI. The study proceeded in iterative cycles of testing, feedback, and refinement. Participants' task was to use the app as they would any other AI chat interface, in at least three different situations which were as diverse as possible. The testing phase lasted between four and five days to give users time to engage with the app more than three times if they wanted to. The design team regularly monitored the incoming data and assisted participants as needed. In general, after the first use when eventual browser permission issues were encountered, the testing phases proceeded without major interruptions. 

After completing the task, participants filled out a feedback survey with Likert-scale questions customized to each version of the probe to according to the main goals addressed by that probe version. Two open-ended questions were included in all feedback surveys: ``How was your experience using this version?'' and ``What are your expectations for the next version?''. The design team then discussed the results and set design goals for the next iteration.

\begin{figure*}
    \centering
    \includegraphics[width=.8\linewidth]{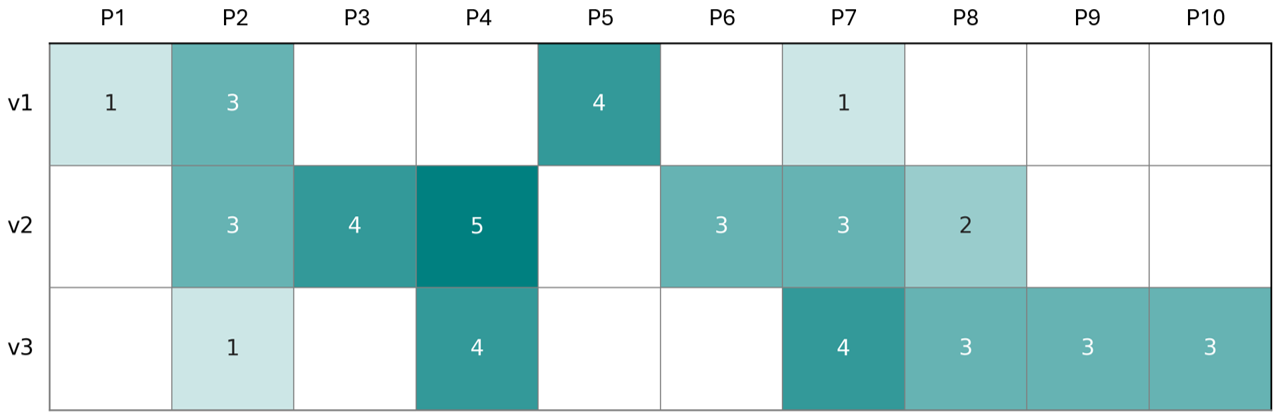}
    \caption{User activity across iterations of the design probe. A total of 47 conversations were documented across 18 unique locations, situated within either virtual or physical environments.}
    \label{fig:user_engagement}
\end{figure*}

To allow maximum time flexibility and reduce the participation barrier, we favored lightweight studies that took at most 30 minutes to complete. As compensation for their engagement, participants were granted two tickets to different raffles after completing a feedback survey. This incentive model aimed to motivate recurrent participation and enable participants to draw comparisons across versions. Feedback surveys always asked whether the participants wanted to be contacted for the next design cycle, offering them a way to withdraw from the study and still use their raffle tickets. Figure~\ref{fig:user_engagement} shows the distribution of user engagement over the entire study.

\subsection{Analysis}

\paragraph{Concurrent Analysis of User Feedback and App Usage}
To ensure timely responses to user feedback during the study, we conducted lightweight analyses in parallel with the design cycles. The design team reviewed open-ended feedback, analyzed stacked bar plots of Likert scale survey responses (Figure~\ref{fig:v1_survey_likert-label}), and examined heatmaps of app usage (Figure~\ref{fig:user_engagement}). These discussions guided the design goals for subsequent iterations. Section~\ref{sec:probe} provides a narrative~\cite{bardzell2016documenting} summarizing the key findings and updated design goals for each cycle.

\paragraph{Affinity Diagramming of User Conversations}
To analyze the range of goals users pursued during the study, we employed affinity diagramming~\cite{gray2014reprioritizing} to identify meaningful clusters. Two researchers (the second and third authors) collaboratively reviewed and organized user prompts into thematic categories. While grouping prompts, the researchers independently assessed whether the probe's first response to the prompt was satisfactory, with agreement determined by an AND operation. This iterative process involved grouping similar prompts based on their goal and required AI functionality. The resulting user goal clusters are presented in Section~\ref{sec:convos}.

\paragraph{Thematic Analysis of User Feedback}
At the end of the study, we revisited surveys and prior discussions in a in-depth reflexive thematic analysis presented in Section~\ref{sec:themes}. Our reflexive thematic analysis followed the six-step process outlined by Braun and Clarke~\cite{braun2006using, braun2024deductive, clarke2013successful}. Three authors participated in the analysis, each bringing a unique perspective to the process: one actively coded the probes, another designed the surveys and collected participant feedback, and the third used the app to gain insights from a user’s point of view. In the familiarization phase, we compiled and organized the data by survey question and participant. The analysts also gained familiarity throughout the study, as they actively discussed user feedback to take design decisions in each version of the probe. Each author independently coded all responses, generating initial codes and applying iterative refinement to their work. The authors compared their codes and collaboratively reached a consensus, producing a unified coding applied to the entire dataset. In the next phase, we developed themes by synthesizing and organizing codes to narrate user experiences and feedback. We reviewed those themes to ensure they are representative of the dataset. We then named the themes to articulate their relevance to our inquiry: applying AI agents to CHAI design grounded on user feedback. The final themes are presented as results in Section~\ref{sec:themes}.

\section{Conversational AI Probe}
\label{sec:probe}
This section presents our initial design rationale~\cite{horner2006effective, bardzell2016documenting} to navigate the challenges of ambiguity and transience. The initial design rationale served as starting point for our investigation, as user feedback and emergency of new design inquiries drove our subsequent iterations~\cite{gaver2022emergence}. We narrate the design process highlighting specific design goals, implementation details, and findings in each version~\cite{bardzell2016documenting}. Finally, we present an annotated portfolio~\cite{bowers2012logic, gaver2012should, bardzell2016documenting} of the final probe artifact in terms of stages, technical components and design reflection. A more detailed account of user feedback can be found in Section~\ref{sec:user_feedback}. Implications of our findings are left for Section~\ref{sec:discussion}.

\subsection{Design Rationale}

Our key strategy to navigate ambiguity and transience in CHAI was to leverage the context of the human-AI conversation and share users' cognitive and motor effort with AI agents.

We anticipated that context would help to narrow the scope of users' potential goals and therefore relevant AI functionalities, important sources of ambiguity. For instance, an image of fridge ingredients at 8 AM suggests the user's goal is to prepare breakfast. A relevant AI function would be generating a breakfast recipe using the identified ingredients. This educated guess based on shared context between the user and AI reduces ambiguity in subsequent conversations. Leveraging multimodal contextual information enables more concise communication between users and AI, which may be preferable to textual or verbal messages in transient interactions. Having access to contextual information from previous user interactions can help designers to disambiguate user goals at design-time and also to identifying patterns of more long-lasting behavior that transcend each transient conversation. This invaluable design resource can help designers to better support specific AI affordances for a given audience's goal.

Agents that share the same context as users can work as helpers that generate goals and prompt recommendations for the user, saving time and effort in a transient conversation. Prompt recommendations by AI agents can work as affordances that connect a potential user goal with an AI functionality, helping users to navigate ambiguity in CHAI. Agents can also support users in refining their goals and prompts supporting cycles of divergent and convergent thinking using minimal user input that require less effort than trial-and-error prompting. From the designers perspective, these agents work as autonomous designers---instructed by a human designer---that are always available at interaction time, and can stand-in for the designer in all  transient interactions by testing ``micro-hypotheses'' of prompts and refining users' goals. Similarly, agents role-playing users can serve as interactive personas for to support designers in early design iterations~\cite{shin2024understanding}, overcoming transience as these agents are perennial and reducing ambiguity for designers through dialog.

We anticipated that a structured workflow, outlining clear steps and responsibilities for both humans and agents, would facilitate effective collaboration while providing a systematic approach to CHAI design. Beginning with the second version, our user-centered workflow included three stages: contextualization, goal formulation, and prompt articulation. In parallel, designers created agents and analyzed user feedback and app usage data within a separate designer-centered workflow.

\subsection{Version 1 - From Context to Prompt}
\begin{figure*}
    \centering
    \includegraphics[width=1\linewidth]{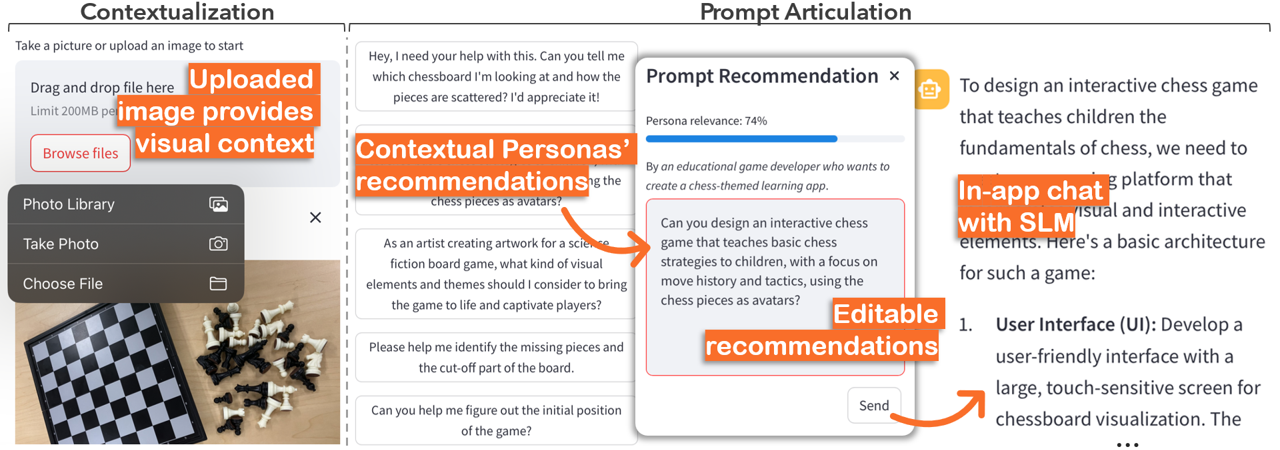}
    \caption{v1: Our main hypothesis was that contextual persona's recommendations would resonate with users due to the shared context provided by the uploaded image. Users could edit a selected recommendation before starting the in-app chat with the SLM that automatically included the uploaded image.}
    \label{fig:v1}
\end{figure*}

\subsubsection{\textbf{Design Goals}}
Following our design rationale, our first design aimed to reduce ambiguity while accommodating transience in CHAI with agent-generated prompts based on visual context. As shown in Figure~\ref{fig:v1}, users provide visual context by uploading an image or taking a photo of their environment, virtual or physical. This input requires minimal user effort, which is convenient in short transient interactions where longer and more effort-demanding forms of input could be impractical. We expected that prompt recommendations based on the visual context would help users to navigate ambiguity by narrowing down the scope of their goals.

Version one (v1) utilized user-provided visual context to generate agents that recommended prompts. These prompts acted as \emph{micro-hypotheses} of affordances, enabling us to design in a transient setting. We expected to align recommendations with users' true goals, to disambiguate AI's functionalities. This design suits transient interactions because recommendations are automatically produced by agents and require no effort from users.

\subsubsection{\textbf{Implementation}}

The main technical component in version 1 was the \textbf{Contextual Persona agents}. Their implementation began with captioning the user’s uploaded image using a multimodal SLM~\cite{abdin2024phi}, followed by embedding the resulting textual description into a 384-dimensional vector using a sentence embedding model~\cite{reimers2019sentence}. This embedded vector was then used to query a large-scale persona hub dataset~\cite{ge2024scaling}, retrieving the three persona descriptions with the highest cosine similarity to the user’s context. Each retrieved persona was animated through role-play prompts, resulting in agents semantically related to the context. These agents generated specific requests for a hypothetical virtual assistant, aimed at fulfilling their inherent needs in relation to the visual context. The system presented these agent-generated requests as prompt recommendations in the form of clickable buttons on the interface. Selecting a button revealed a pop-up containing the persona’s description and a bar chart visualizing its relevance, calculated by mapping cosine similarity values into percentages. Additionally, users were able to edit the generated prompts before engaging in an in-app chat with the SLM, which was pre-loaded with the context image.

\subsubsection{\textbf{Reflection}}

Users highlighted significant gaps in the alignment of suggested prompts with their intended goals. Many found the prompts overly specific to the Contextual Persona agents or irrelevant due to mistakes in visual analysis, such as confusing an orthotic for a cast. For these reasons, the practical utility was limited, especially when users had clear goals upfront. Users reported that the workflow was too rigid and sometimes prevented them from making progress toward their true goal. Perhaps because of the low satisfaction with the initial prompt recommendations, some users deleted the suggested text and wrote their own prompts from scratch.

Despite these issues, users saw potential value in the system for straightforward and open-ended goals. The unexpected ideas generated by Contextual Persona agents enabled interesting exploratory directions. P1 reported that reading the persona description helped them understand that prompt was a plausible recommendation coming from that agent.

Revisiting our initial design rationale, we interpreted these results as evidence that the range of potential user goals was too large to find the user's true goal, even with visual context and a large scale persona dataset. Analyzing usage logs, we noticed that many recommendations were indeed semantically related to the context, but unrelated to the prompt users used in the in-app chat, suggesting a goal mismatch. For example, P5 provided a screenshot of a puzzle web game described as \textit{``a grid of tiles with various words on them,''}. The system generated a Contextual Persona agent who was \textit{``a graphic designer,''} and recommended prompts including \textit{``What kind of error message design ideas do you have for a software application?''}. P5 edited the prompt to \textit{``I need to group these 16 words into 4 groups for a puzzle.''}

\subsection{Version 2 - Decoupling Goals from Prompts and Personalization}

\begin{figure*}
    \centering
    \includegraphics[width=1\linewidth]{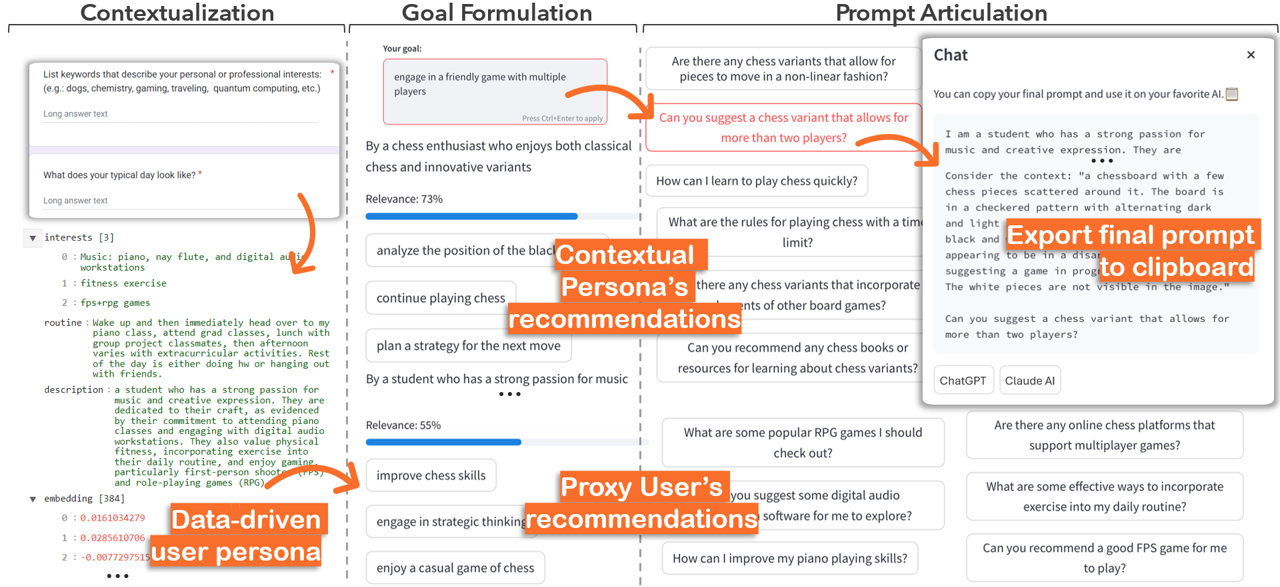}
    \caption{v2: In response to users' desire for more personalized recommendations, version 2 introduced recommendations from a Proxy User agent generated based on users' routines and interests. In this version we decoupled the goal formulation from the prompt articulation, providing goal recommendations that served as input for the prompts. To leverage the power of more advanced models, we replaced the in-app chat with the ability to export the prompt to a third-party AI chat application.}
    \label{fig:v2}
\end{figure*}

\subsubsection{\textbf{Design Goals}}
Version two (v2) aimed to improve the relevance of recommendations by adding personal context and separating goals from prompts.
We introduced a Proxy User agent, who role-played the user's persona elaborated with pre-survey data, in order to enable more personalized recommendations. This version kept Context Persona agents to support creative exploration. Instead of directly recommending prompts based on context, this version first presents the user a goal formulation stage, allowing them to refine goals before articulating prompts.

\subsubsection{\textbf{Implementation}}
\textbf{Proxy Users} are agents generated by role-playing prompts based on users’ personas. These personas were created using data from pre-study surveys, which included descriptions of users’ routines and keywords reflecting their professional and personal interests. We expected that Proxy User's recommendations would be more aligned with the users' true intentions.

The new goal formulation stage consisted of first asking agents to list their needs and goals given the presented context. Users selected a goal from these lists and made edits when necessary, thus guiding the recommendation process with user feedback and improving alignment with the actual range of user goals. This stage is similar to the chain-of-thought reasoning technique, which improves LLMs performance in many tasks~\cite{wei2022chain, stechly2024chain}, while interleaving the links of the chain with user input.

The selected goal was then assigned to the same Proxy User and Context Persona agents from the previous stage who generated prompt recommendations by making a request to a hypothetical virtual agent. Similarly, users could edit the final prompt before proceeding to the chat.

Acknowledging the limitations of our SLM-based probe compared to current AI chat products, we chose not to include an in-app chat in this version. Instead, we allowed users to export the final \emph{super prompt}, containing contextual information and goal statement, to the clipboard. Users could use link buttons to take them to their favorite AI chat application where they could paste the final prompt and continue the conversation there.

\subsubsection{\textbf{Reflection}}

Participants welcomed the possibility to clarify their goals before receiving prompt recommendations, suggesting that a separate goal formulation stage can stimulate reflection before engaging in trial-and-error prompting. \textit{``This step of first identifying goals helps me explore more potential ideas''}---Gold. We consider the goal formulation stage can be further explored as a way to mitigate the ``Intentionality Gap''---difficulty to assess AI's output due to the lack of clear intentions~\cite{subramonyam2024bridging}. Participants P7 and P4 wanted goals and prompts to be more clearly differentiated at the interface level, indicating this segregation was already happening at their mental model.

Users appreciated having personalized prompts derived from their own personas, though the perceived relevance of the suggestions varied greatly. Some participants reported their persona description was accurate but not parts of it were relevant to the particular conversation, which caused irrelevant recommendations when the User Proxy agent fixated on these parts. For example, P2's persona included ``product design'' and given the context of a wine menu, the Proxy User recommended ``organize the wine list by color categories'' when the user final prompt was ``Can you suggest a wine from the list that pairs well with a seafood dish?''.

Regarding the prompt articulation stage, P2 noticed that the recommended prompts seemed to restate user goals verbatim rather than expanding on them, suggesting more value could be added to the prompt articulation stage.

\subsection{Version 3 - More User Control}

\begin{figure*}
    \centering
    \includegraphics[width=1\linewidth]{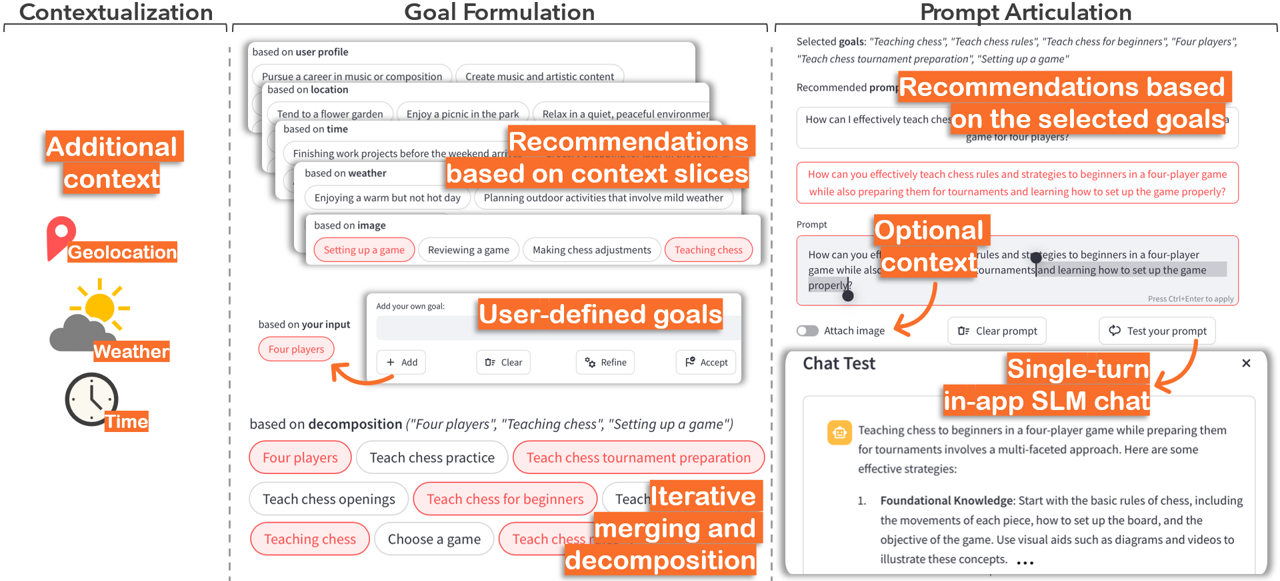}
    \caption{v3: the main goal of the probe became to support goal formulation. We introduced additional context while keeping recommendations focus on one slice of context at a time. Users could also directly input goals not captured in the recommendations. After selecting relevant goals, users could refine them using iterative merging and decomposition of the selected goals. Finally, the system generated prompts based on the selected goals and users could iteratively test their prompts with the in-app SLM chat.}
    \label{fig:v3}
\end{figure*}

\subsubsection{\textbf{Design Goals}}
In this version, we aimed to improve goal recommendation relevance by leveraging iterative user input to refine goals and filter context. We expected that goal formulation could help to navigate transience as well, as a clear goal can have a long-lasting meaning that spans multiple conversations and creates design opportunities that transcend the scope of a single conversation.

\subsubsection{\textbf{Implementation}}
Building on findings from version 2, we adopted a more cautious approach to using context, as we observed that irrelevant context can amplify ambiguity. In this version, the Proxy User and Contextual Persona agents processed one type of contextual information at a time (e.g., visual context, location), instead of using the entire context at once. Goal recommendations were grouped by context type, rather than by agent as in previous versions. Users could select multiple goal recommendations or input their own goals if none of the recommendations reflected their actual goal.

Users could refine their selected goals during a nested workflow within the goal formulation stage. In this process, they collaborated with a \textbf{Goal Refinement agent}, which combined user-selected goal recommendations with user-defined goals and then performed a decomposition operation to break them into smaller, actionable sub-goals. Users could update their selection and iteratively repeat the goal refinement process until they were satisfied. This iterative refinement merged and subdivided goals, enabling users to choose smaller actionable goals.

The resulting goals were set to agents in the same prompt articulation method previously described. Finally, users could edit and test prompts in a single-turn in-app chat, enabling a quick quality check. This allowed users to refine prompts before exporting them to a third-party AI chat application.

\subsubsection{\textbf{Reflection}}
The user feedback highlights a mix of positive comments and constructive suggestions about the goal refinement and recommendation processes. Users reacted positively to the presentation of recommendations in groups based on context type. Some appreciated the categorization stating that it helped them navigate and refine goals more efficiently because it reduced the need to read through all suggestions, allowing them to focus on categories aligned with their needs. Regarding the content of the recommendations generated from different context types, some users found that categories such as time, provided irrelevant suggestions, especially when not aligned with the provided image or other input. Curiously, another participant who used the app on the new year's eve highlighted a time-based recommendation to ``prepare for a celebration with friends'' as being surprising relevant. Others noted that they primarily interacted with recommendations based on user persona and image while the other groups were less useful. This suggests that the relevance of context types fluctuates and giving users the power to combine and segment context types may enhance recommendations.

Users appreciate the ability to refine goals by selecting from suggestions and manually adding their own, which facilitates creativity and alignment with their needs without overwhelming them with options. Users suggested emphasizing the ability to input goals manually, perhaps even before generating recommendations in future versions.

As designers, we found with version three that context granularity is an important CHAI design factor. User input seems to be a promising pathway to filter context at a meaningful granularity. Users demonstrated engagement with the iterative goal refinement process, which can be further enhanced in future designs by allowing user to mix and match pieces of contextual information. Moreover, it became clear that recommendations generated upfront are less effective and engaging users in a incremental process where both context information and user input is combined is a more promising avenue for navigating ambiguity in transient conversational interactions with AI.


\subsection{Version 4 - Designers-in-the-Loop}

\begin{figure*}
    \centering
    \includegraphics[width=.4\linewidth]{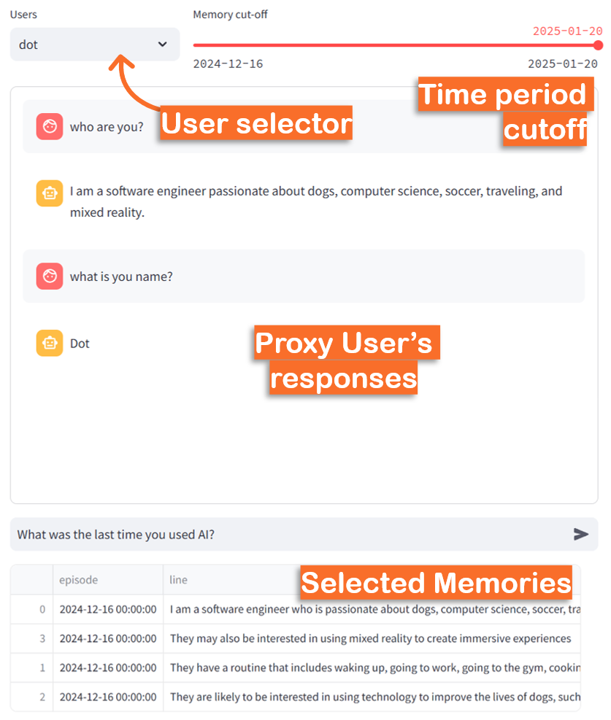}
    \caption{v4: The designer-facing interface enables selecting a specific user to create a Proxy User agent for conversational interactions grounded in app usage data and user survey responses. A slider allows designers to set a cutoff date, restricting the agent’s memory to data available up to a specific release.}
    \label{fig:v4}
\end{figure*}

\subsubsection{\textbf{Design Goals}}
Transience in CHAI significantly reduces opportunities for designers to engage with users, a challenge we experienced firsthand. It was difficult to capture the breadth of users’ goals and prompts within our probe, and our design iterations relied primarily on static app usage logs and survey responses. To address this limitation, we introduced the Interactive Proxy User in our final version. By offering a more immersive and dynamic way to engage with user data, this feature aims to help designers develop a deeper, more nuanced understanding of user experiences.

\subsubsection{\textbf{Implementation}}
Building on our previous implementation of Proxy Users and inspired by the interactive personas proposed by Shin et al.~\cite{shin2024understanding}, we developed a designer-facing interface that lets us select a particular user and converse with a Proxy User agent role-playing that user. This agent is enriched with memories of past app interactions, which are parsed, semantically embedded, and indexed to allow cosine-similarity lookups. A RAG approach selects relevant logs based on prompts, enabling an in-depth exploration of how users engage with the app through a CHAI interface. Designers can also adjust a slider to limit the agent’s memory to specific release dates. Figure~\ref{fig:v4} shows the design-facing interface.

\subsubsection{\textbf{Reflection}}
Interacting with Proxy Users through conversations grounded in app usage data proved to be a feasible and valuable approach, complementing traditional methods like surveys and static app log analysis. As authors and designers, we tested the designer-facing interface and found it functioned effectively as a conversational search engine, helping identify critical information within the app log documents. We observed that Proxy Users role-playing participants with the most extensive data collected were the most useful. This suggests that the approach is particularly insightful when there is a rich dataset to draw from, highlighting the importance of comprehensive user data for effective role-playing and meaningful interactions.

\section{User Feedback}
\label{sec:user_feedback}

In this section we discuss user feedback from two different perspectives: first looking at the thematic analysis of user interaction and experience, and then understanding user conversations when using the probe.

\subsection{Themes}
\label{sec:themes}
Based on the analysis of participants responses and researcher-coded insights, we identified three overarching themes. Each theme encapsulates specific aspects of participants feedback and highlights critical areas of focus for improving conversational AI workflows. Below, we detail these themes.

\subsubsection{\textbf{Divergent and Convergent Thinking}}
\paragraph{\textit{Experimentation and Idea Exploration.}}
Participants frequently described using the system as a ``creative sandbox'' for trying new ideas. In these moments, they explored multiple directions or asked for varied suggestions---reflecting a divergent thinking approach. Some found it particularly helpful when the system offered prompts they would not have considered on their own. For example, P5 said, ``It was interesting, [...] I only used it in an exploratory way.'' or P4 stated, ``I think it allowed you to embody personas that were kind of out-of-the-box. That was very interesting to see.''

\paragraph{\textit{Transitioning to Focus and Refinement.}}
Once participants had a clearer sense of the end goal, they shifted from open-ended exploration (divergent) to more targeted refinement (convergent). For instance, P9 remarked, ``it helped refine and clarify the goals I had''. Many explained that while they initially wanted an array of suggestions, they gradually needed to merge, discard, or reword prompts to form a single cohesive solution or final goal.  

\paragraph{\textit{Influence of Uncertain or Incomplete Information.}}
Some participants felt stuck in exploratory mode when confronted with ambiguous or unclear prompts, especially when using v1. Because the system’s suggestions sometimes lacked context or leaned on assumptions the user did not share, participants circled back to broader brainstorming. For example after testing version 1 of the system, P4 shared, ``Some goals and prompts can go in many directions if there is need for creativity.'' This exploration and refinement underscores how the system can both foster creativity and demand extra user intervention when suggestions are off-target.

\subsubsection{\textbf{Experience with Recommendations}}
\paragraph{\textit{Relevance and Alignment.}}
Participants generally praised prompts that aligned well with their goals or personal context, finding that they saved time and sparked fresh ideas. Although not especially frequent, there were a few instances where participants noted that prompts lacked relevance or were based on misguided assumptions. In those less common situations, some reported frustration or questioned the system’s utility. Overall, however, many felt that the benefits of well-aligned prompts outweighed occasional mismatches.

\paragraph{\textit{Personalization and Persona-based Prompts.}}
A notable feature for many was the system’s attempt to personalize recommendations---whether by drawing on user-specific details or by using ``persona-inspired'' suggestions. When the persona or contextual details truly reflected the participant’s background or intentions, the recommendations felt purposeful and engaging. P7 commented, ``It suggests prompts that I have never thought of.'' By contrast, any mismatch (e.g., suggestions rooted in irrelevant interests) generated confusion or broke the sense of personalization. Participants described these mismatches and called for more nuanced ways to tailor prompts.

\paragraph{\textit{Balancing Novelty and Practicality.}}
While participants enjoyed discovering novel ideas through automated prompts, they also wanted suggestions to feel applicable to their current task. Some viewed wild or unconventional prompts to be fun but less helpful in specific goal accomplishment. Others saw value in ``push-your-boundaries'' suggestions, as long as the system provided enough rationale for why those suggestions might be relevant.

\subsubsection{\textbf{Control and Automation}}
\paragraph{User-Driven Versus System-Driven Prompting.}
Different participants had different preferences for how much they wanted to control the prompt-generation process. After testing version 2 of the system, P3 mentioned ``[it] would be nice to add multiple recommendations to my goal, and edit the context + personal background''. We added these features in version 3 and participants relished the autonomy of manually customizing or editing suggestions. Some of them still appreciated automated features that quickly consolidated ideas and proposed next steps, especially when under time constraints.

\subsection{Conversations}
\label{sec:convos}
\begin{figure*}
    \centering
    \includegraphics[width=.9\linewidth]{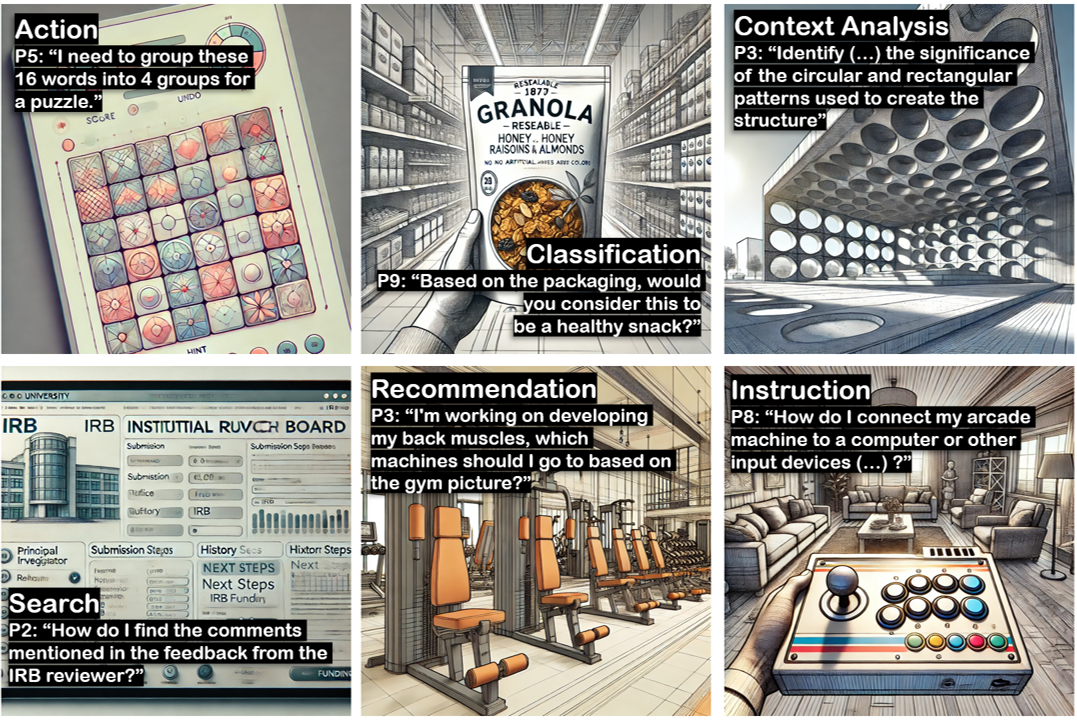}
    \caption{Users interacted with the probe using prompts we categorized in action, classification, context analysis, search, recommendation, or instruction. These prompts were either context-free or context-bound, when they required additional context provided by the user beyond the prompt itself.}
    \label{fig:prompts}
\end{figure*}

Users engaged with the probe a total of 47 times in 18 different locations pursuing goals in both physical and virtual environments, such as using exercise equipment for strength training or finding information on a webpage. Two authors categorized these goals using affinity diagramming~\cite{gray2014reprioritizing} to create semantic clusters. Through this process, we identified two broader categories: context-free prompts and context-bound prompts. When properly formed, context-free prompts could be answered by the virtual agent using only information from the prompts. In contrast, context-bound prompts required the virtual assistant to refer to the additional information provided by the user such as image or location. Within these major categories, our analysis identified six other categories as seem in Figure~\ref{fig:prompts}.

Action prompts delegated the resolution of a problem to the virtual agent, e.g., ``I need a larger, more detailed image to help me accurately describe and rate the wines listed in your photo.''---P2. Classification prompts were requests to assign a label to an object or piece of information, e.g., ``What cuisine does this restaurant serve?''---P10. Context analysis prompts demanded the virtual agent to interpret the context or a piece of information explicit in the prompt, e.g., ``Can you tell me more about the details provided on the (puzzle trademark) set box (...)?''---P2. Search prompts requested an information lookup, e.g., ``What are the macro-nutrient breakdowns for a bowl of mixed vegetables and tofu?''---P4. Recommendation prompts requested the virtual assistant to make a suggestion based on some prior, e.g., ``What are some of the best places to take photos of the city at night?''---P6. Finally, instruction prompts asked for guidance often in the form of a step-by-step procedure, e.g., ``How can I make sure my dog is comfortable and happy during a photo shoot?''---P4.


\section{Discussion}
\label{sec:discussion}

In our RtD study, we departed from the rationale that adding context to conversations and distributing cognitive and motor tasks between users and AI agents in a structured workflow could help mitigate the challenges of ambiguity and transience in CHAI. Over four iterative design cycles with our probe, we observed that these strategies do support both designers and users in dealing with these aspects of CHAI. However, our findings also indicate the importance of careful design to avoid drawbacks such as excessive automation and addition of irrelevant context. This section discusses these nuances as well as implications of our findings in other areas of design.

\subsection{Goal Formulation as Outcome}

The effectiveness of human-AI conversations hinges on the clarity of the user's goals. A key contribution of CHAI design lies in supporting users' cognitive processes for refining and articulating their goals. Unlike traditional direct manipulation interfaces that focus on task execution by providing affordances aligned with users' mental models, CHAI shifts the burden of goal specification to users while automating task execution. This shift places greater emphasis on designing for goal formulation rather than execution support~\cite{subramonyam2024bridging}.

In the first version of our probe, prompt recommendations often did not satisfy users and demanded re-working the prompts from scratch through trial-and-error prompting with the in-app chat. User feedback pointed that prompts were coherent with their context but reflected goals that were not relevant to them. Users reacted much better when the goal formulation and prompt articulation stages were decoupled, in special in the last version, that introduced a workflow to refine recommended goals and user-defined goals through merging and decomposition operations.

Compared to purely conversational metaphors, this design could reduce motor effort in goal formulation by minimizing extensive prompt writing. Instead, direct manipulation~\cite{hutchins1985direct} elements, such as toggles for selecting AI-generated goals recommendations (Figure~\ref{fig:v3}), streamline the process. Cognitive effort was shared with Contextual Persona and User Proxy agents, which users found valuable for divergent thinking, and Goal Refinement agents, which supported convergent thinking through goal merging and decomposition.

\subsection{Side-Effects of Contextualization}

Our initial design rationale suggested that enabling users to share context could narrow the scope of relevant goals and AI functionalities, thereby reducing ambiguity. However, a dose-effect gradient must be considered. Both too much and too little context can lead to undesirable outcomes. Including excessive contextual dimensions may cause the AI system to fixate on irrelevant details, diverging from the user's true goal and increasing ambiguity rather than reducing it. For example, P2 said, ``Yes, I'm someone with a background in product design but that is irrelevant when I am choosing wine.'', suggesting that their User Proxy persona was accurate but lead to recommendations not aligned to their actual goals.

Identifying which aspects of context are relevant is challenging, and our observations highlight the importance of user input in this process. Therefore, contextualization should be paired with user choice to filter and prioritize relevant dimensions within an iterative workflow. Clearly explaining which context modality was used to generate the recommendation and allowing users to steer the recommendations by providing feedback proved to be a more effective strategy. We recommend CHAI designers to prioritize features that support users in expressing their goals over collecting too much contextual data, which might not be helpful to proactively search the range of user goals.


Sharing contextual data often involves a degree of privacy compromise, and this trade-off should be made explicit to users. For example, sharing details about one’s routine may not always justify the conversational benefits. Allowing users to explicitly attach or detach specific aspects of context empowers them to decide what to share based on relevance to their use case, rather than requiring them to share all data by default.

With that said, our study was intentionally broad in term of use case to capture the ambiguity of scope in general CHAI. Contextual data might be more effective for anticipating users' intentions in more constrained scenarios with limited scope of intentions such as physical task support~\cite{arakawa2024prism}.

\subsection{Personas as Anthropomorphic Signifiers of Agents Affordances}

Users reported that presenting recommendations alongside the description of the agent's persona that generated them enhanced their ability to relate to the prompt and goal. This approach promoted cognitive engagement by helping users recognize alignment with their objectives or by encouraging alternative perspectives, fostering divergent thinking. As a result, users could better explore the range of AI functionalities.

Personas functioned as an explainability feature, aiding users in contextualizing and understanding how recommendations were generated. In the final version of our probe, personas were also used to communicate AI affordances. Users noted that these personas provided insight into the system's reasoning, serving both as explanations and signifiers of the system's state in response to user context.

Despite limitations in recommendation quality, leveraging personas to signify AI affordances shows promise. Users often attribute human-like characteristics to conversational systems, and embracing this as a mental model—rather than treating it as a drawback—may enhance usability and engagement, as observed in our study.

\subsection{Conversational AI Probes as Needfinding Machines}

Our study, in particular the last version of the probe, indicates that a conversational AI probe can serve as a powerful ``Needfinding Machine,'' revealing both explicit and latent user needs~\cite{martelaro2017needfinding}. By supporting users to articulate well-formed intentions and then observing whether the underlining AI model can fulfill those intentions, we establish an iterative affordance discovery process.

By analyzing conversations that fulfilled user goals, designers can identify recurring demands that overcome transience to become a long-lasting need. In this case, prolonged user engagement in user-centered or participatory design becomes a feasible approach, suitable to transition from a low-fi conversational prototype to a tailored solution. Moreover, conversations fulfilled by the conversational AI probe indicate that the underlining AI model has a relevant functionality, offering technical opportunities for a novel design. Conversations where the probe fails to meet the user's goal are also valuable design opportunities. In this case, designers can identify the limitations of current design and enabling AI model that require deeper user research or a new design exploration.

Though we focus on conversational AI, this method could be applicable to other design domains, where a conversational AI probe would act as a low-fidelity prototype for more specialized systems yet to be designed. Deploying a conversational system in the wild can generate a continuous feed of design insights. Ultimately, this approach leverages the flexibility of conversational AI as a rapid exploration method while providing meaningful data to guide subsequent high-fidelity prototyping and product development.

\section{Limitations and Future Work}

This study spanned four design cycles over one month, presenting only the most feasible and compelling designs given our constraints. We encourage researchers to explore alternative workflows, recommendation techniques, and tailored persona datasets, as the persona hub used here was not optimized for generating recommendations~\cite{ge2024scaling}. The study used Phi-3.5 Vision~\cite{abdin2024phi}, a small vision language model, potentially limiting performance in reasoning tasks when compared with a language-only equivalent. Future work should consider larger or specialized models, including Phi-4, which was released during our study~\cite{abdin2024phi4}. Longer-term studies could better capture recurring goals and integrate the app into users' routines, providing deeper insights into transient and persistent functionalities. While our diverse sample included participants of various ages and who declared to have or previously had a disability, studies with a focus on accessibility are also needed.

\section{Conclusion}
We presented an RtD study exploring agentic workflows incorporating humans-in-the-loop and contextual data to help users and designers navigate ambiguity and transience in conversational human-AI interaction. Across four design cycles with ten participants over one month, our findings highlight the benefits and challenges of contextualization in addressing goal ambiguity and AI functionality, emphasizing the need for careful application. Decoupling goal formulation from prompt articulation enabled the use of agents specialized in these distinct stages. Goal formulation emerged as a critical phase in CHAI, where users benefited from collaboration with Goal Refinement agents and recommendations from Contextual Persona and User Proxy agents. These agents allowed users to test micro-hypotheses for goals and prompts dynamically, addressing the challenges of transience in interaction design. Additionally, enabling designers to interact with User Proxy agents that accessed real app usage data provided a novel approach to overcoming transience in CHAI design. These findings offer broader implications for design, particularly in leveraging conversational AI probes as low-fidelity prototypes for tailored systems yet to be developed.


\bibliographystyle{ACM-Reference-Format}
\bibliography{
    bib/main,
    bib/methods/thematic_analysis,
    bib/methods/design_approaches,
    bib/methods/diary,
    bib/methods/rtd,
    bib/ai,
    bib/chai,
    bib/cui,
    bib/design_with_ai,
}
\end{document}